\documentclass[12pt]{article}
\usepackage{amsmath}
\usepackage{graphicx}
\usepackage{float}
\newcommand{\be}{\begin{equation}}
\newcommand{\ee}{\end{equation}}
\newcommand{\ba}{\begin{eqnarray}}
\newcommand{\ea}{\end{eqnarray}}

\begin{document}

\title{{\bf Possibilities for Probabilities}
\thanks{Alberta-Thy-8-22}}
\author{
Don N. Page
\thanks{Internet address:
profdonpage@gmail.com}
\\
Theoretical Physics Institute\\
Department of Physics\\
4-183 CCIS\\
University of Alberta\\
Edmonton, Alberta T6G 2E1\\
Canada
}
\date{2022 September 14}

\maketitle

\begin{abstract}

In ordinary situations involving a small part of the universe, Born's rule seems to work well for calculating probabilities of observations in quantum theory.  However, there are a number of reasons for believing that it is not adequate for many cosmological purposes.  Here a number of possible generalizations of Born's rule are discussed, explaining why they are consistent with the present statistical support for Born's rule in ordinary situations but can help solve various cosmological problems.

\end{abstract}


\section{Introduction}

Bare quantum theory describes the universe in terms of quantum states, operators, expectation values, wavefunctions, amplitudes, path integrals, etc.\ but does not by itself connect these to observations, so bare quantum theory is not directly testable.  To make comparison with observations, Max Born \cite{Born} suggested that the wavefunction gives {\it probabilities}.  Later it was proposed that the probabilities are given by the absolute squares of amplitudes, and these can be taken to be expectation values of projection operators, so in this paper I shall take the mathematization of Born's rule to be the idea that the probabilities of observations are given by the expectation values of projection operators.

\newpage

\section{Problems with Born's Rule}

Born's rule seems to work well empirically in the small part of the universe in which our human observations have been carried out (counting only the regions where the human observations have been made, on or near the earth, not the vast part of the observable universe from which the information propagated before being observed by humans).  However, there are conceptual reasons for doubting that Born's rule applies over the whole universe or that it is even adequate for explaining the probabilities of our observations from a fundamental theory for the universe when one takes into account the possibility that our observations could occur in many places across a vast universe \cite{Page:2008ns,Page:2009qe,Page:2009mb,Page:2010bj}.  For example, if there are two identical copies of an observer who each measure the vertical component of the spin of an electron, and if the electron measured by one copy of the observer has spin up and the electron measured by the other copy of the observer has spin down, and if the observer is uncertain which copy he or she is, the probability of observing spin up will be between 0 and 1, which is not the result that would be given by the expectation value of a projection operator that is chosen independently of the quantum state.\\

Other observations that at least na\"{\i}vely seem difficult to explain by Born's rule include our observations that we are humans rather than ants, despite the much greater number of ants on earth, and our ordered observations that do not seem consistent with typical Boltzmann brain observations that appear to dominate observations in many simple models for the universe that include an application of Born's rule.  As we shall see, modifications of Born's rule can explain the probabilities of observations by copies of observers, the fact that we do not observe ourselves to be ants, and the fact that we apparently do not observe ourselves to be Boltzmann brains.  However, it is not yet known what the correct modifications are, so there remain many possibilities for probabilities of observations.

\section{Sensible Quantum Mechanics}

Here I shall not give an exhaustive list of possible replacements for Born's rule but just describe a few of the basic possibilities.  The simplest set of possibilities seem to be that the relative probabilities of observations are given by the expectation values of positive `awareness operators' $A_j$, one for each observation $O_j$, that are not projection operators \cite{Page:1995dc,Page:1995kw,Page:2001ba,Page:2011sr,Page:2014aqa}.  These possibilities give a framework that I have called Sensible Quantum Mechanics (SQM) or Mindless Sensationalism (MS).  Then if angular brackets denote the expectation value of an operator in the quantum state of the universe, the probability of the observation $O_j$ is
\ba
P(O_j) = p_j/\sum_k p_k,
\label{p-obs}
\ea
where the unnormalized relative probability of the observation $O_j$ is
\ba
p_j = \langle A_j \rangle,
\label{a-exp}
\ea
the expectation value of the awareness operator $A_j$ in the quantum state of the universe.

Sensible Quantum Mechanics (SQM) is a completely deterministic framework (once the quantum state and the awareness operators are selected; they are not logically determined by the framework) in which the quantum state (in the Heisenberg picture) never changes or collapses, so there is no true randomness and no true probabilities in the sense of propensities for potentialities to become actualities.  The framework also does not imply any randomness for observations.  What I am calling the ``probability of an observation'' in SQM is actually a normalized measure for it, giving in some sense how much it exists, but all observations with nonzero measure do have some measure of existence, determined by the quantum state and the awareness operators.  It might be helpful to think of the observation $O_j$ as being selected at random with the probability given by the measure $P(O_j)$, but SQM denies that there is any such random selection; instead, all observations $O_j$ with positive measure actually occur, but just with different measures given by $P(O_j)$.  In this sense SQM is a ``many-perceptions'' framework rather than a ``many-worlds'' framework as the Everett many-worlds version of quantum theory is, though SQM shares with the Everett version the assumption that the quantum state is objectively real and never collapses.\\

One should also note that in the most basic form of Sensible Quantum Mechanics that I have proposed, the observations are conscious perceptions or sentient experiences, each being all that one consciously experiences or is consciously aware of at once.  Since conscious perceptions are directly what we experience, they seem to be the best candidates for what has measures that are analogous to probabilities.  As basic elements of the assumed ontology, conscious perceptions cannot be defined in terms of any other more basic elements, but only pointed to by a description such as the one above.  Conscious perceptions are also not logically implied to exist purely as a consequence of the quantum state; the existence of awareness operators with nonzero expectation values is logically independent of the quantum state itself.\\

With measures for existence confined to conscious perceptions, one does not need to propose that any other aspect of the world has anything like probabilities, rather than just amplitudes and expectation values of operators.  Furthermore, the normalized measure for an observation that is a conscious perception can be used as if it were an objective likelihood of the theory giving this measure in a Bayesian analysis, to be combined with the inevitably subjective prior probabilities assigned to different theories to give the posterior probabilities of the theories for testing them against the observation \cite{Page:2014aqa}.\\

If all the awareness operators formed a complete set of orthonormal projection operators $P_j$ summing to the identity operator, one would have Born's rule.  Indeed, in this case $\sum_k p_k = 1$, so that then one would simply have $P(O_j) = \langle P_j \rangle$.  However, there are many problems with this, such as the ones raised in \cite{Page:2008ns,Page:2009qe,Page:2009mb,Page:2010bj} for identical copies of observers, but also others such as the fact that one would not expect two different observers to have the probabilities of their observations given by orthogonal projection operators, or that all the orthonormal projection operators in any complete set to correspond to observations.  Thus there are many reasons for supposing that if the relative probabilities $p_j$ of observations are given by expectation values of positive operators $A_j$, that these operators would not be an orthonormal set of projection operators $P_j$ but might rather each be weighted sums or integrals of projection operators, or something else similar.\\

In this paper I shall focus mainly on the SQM framework, in which the relative probabilities $p_j$ are expectation values of positive operators $A_j$ and hence are linear functionals of the quantum state.  However, I shall also consider going beyond SQM to nonlinear rules for getting from the quantum state to the probabilities of observations, such as the possibility that each relative probability is a non-unit power of the expectation value of the corresponding positive operator, $p_j = \langle A_j \rangle^s$ with a positive exponent $s\neq 1$.  Although one cannot observationally clearly distinguish SQM theories which have $s=1$ from non-SQM theories with $s$ not equal to 1 but close to 1, SQM theories with $s=1$, and hence with a linear relation between the quantum state and the relative probabilities $p_j = \langle A_j \rangle$, seem to be the simplest possibilities, so I would personally assign them the highest prior probabilities.

\section{What Are the Awareness Operators?}

The main open question in the Sensible Quantum Mechanics framework, in which the relative probabilities (actually measures) of observations are the expectation values of positive awareness operators $A_j$ (one corresponding to each observation), is what these awareness operators are.  This SQM framework does not assume that there is a spacetime, so it could be valid even in formulations of quantum gravity in which spacetime is not fundamental.  However, if the part of the quantum state that gives the dominant contribution to the expectation values of the awareness operators can lead to the approximation of a effective spacetime or of a quantum superposition of spacetimes, one idea is that the contribution to the expectation value of an awareness operator from each spacetime in the superposition is the expectation value of the existence of this spacetime multiplied by a sum or integral over the spacetime of a localized projection operator for each spacetime region.  (This assumes that each awareness operator $A_j$ gives a negligible matrix element between two quantum states corresponding to two different spacetimes.)\\

For example, if a certain brain state leads to a certain observation, then one might think that a sum or integral over spacetime regions of the localized projection operator for that brain state would be its contribution in that spacetime to the expectation value of the awareness operator and hence to the probability for that observation.  (Of course, many different brain states could contribute to the probability of the observation even in one region, as well as rotations and boosts of these brain states, the latter posing a potential problem for getting a finite sum or integral because of the noncompactness of the Lorentz group.)\\

\section{Weighting for the Integral over Each Spacetime}

If in the spacetime approximation one does assume that each spacetime in a quantum superposition of spacetimes contributes to an awareness operator by approximately the sum or integral over that spacetime of a localized projection operator, not only does one need to know what these localized projection operators are, but also what the weights are for the sum or integral over the spacetime.  It would be simplest to take uniform weights, but this seems likely to lead to divergences if spacetime is infinite or has an unbounded expectation value for its 4-volume.\\  

One might try to regularize this infinity by imposing a finite cutoff on the spacetime and then taking the limit of the normalized probabilities when the cutoff is taken to infinity.  However, projection operators are positive operators, and localized projection operators generically have positive expectation values in nonsingular quantum states (including the vacuum), so if the universe expands to become asymptotically empty, the dominant contribution to the expectation values of the awareness operators will be from these positive expectation values of the localized projection operators in the asymptotically empty spacetime, which correspond to Boltzmann brain observations \cite{DKS,Albrecht:2004,Page:2008a,Page:2006ys,Page:2008b,Page:2017qdu}.  Therefore, the regularization of a uniformly weighted integral over spacetime would seem to lead to domination by Boltzmann brain observations.  Surely almost all Boltzmann brain observations would be much more disordered than ours are, so our ordered observations are almost certainly strong statistical evidence against this Boltzmann brain domination.\\

Therefore, to avoid Boltzmann brain domination in the sum or integral of localized projection operators over an asymptotically empty spacetime, it seems necessary to choose weight factors that give convergent integrals that are not dominated by the asymptotically empty regions where almost all observations would be by Boltzmann brains.  For spacetimes with preferred spatial hypersurfaces (e.g., each at some proper time from a big bang or bounce minimal hypersurface) that each have finite 3-volume (though perhaps tending to infinity asymptotically with time), one simple procedure for greatly ameliorating the divergent integrals over spacetime is to divide the contribution over each of these preferred spatial hypersurfaces by the 3-volume $V$, thus taking the contribution at each time
to be the volume average of the expectation value of the localized projection operator \cite{Page:2008zh,Page:2011gq}.  This will make the contribution of each of the preferred spatial hypersurfaces finite, but since this contribution seems likely to go to a constant at late times if the universe becomes asymptotically empty with an asymptotically constant spacetime density of contributions to Boltzmann brain observations, the sum or integral over times would diverge if the universe lasts forever.  Therefore, I have also proposed Agnesi weighting \cite{Page:2010re} to solve this problem, replacing the integral over $dt$ by an integral over $dt/(1+t^2)$, where $t$ is measured in Planck units along the longest timelike geodesic from the big bang or from a globally minimum hypersurface of a bounce to the location of the localized projection operator.  This is admittedly {\it ad hoc}, so a more elegant formulation should be found (see e.g.\ \cite{Page:2014eoa}), but at least Agnesi weighting combined with volume averaging renders the integral over spacetime of the localized projection operators finite and solves the Boltzmann brain problem.\\

Note that the proposed variation of the weight factors with the location within the spacetime is diffeomorphism invariant and hence coordinate independent.  The particular proposal I have made does depend on the existence of a preferred hypersurface, such as the minimal 3-volume Cauchy surface with zero expansion (a bounce), or the limit sequence of Cauchy surfaces approaching an initial singularity (a big bang), and then, at each event away from this `initial' Cauchy surface, on the maximal proper time of any causal curve from the `initial' hypersurface, and on the 3-volume of the Cauchy hypersurface with that same maximal proper time from the `initial' hypersurface.  If one takes, at each event within the spacetime, the maximal proper time from the `initial' hypersurface to that event to be the `age' of the universe at that event, one can simply say that the weighting I have proposed depends on the age and on the volume of the universe at that age, both of which are coordinate-invariant quantities.\\

Of course, the variation of the weight factors with the location might depend on coordinate-invariant quantities that are different from the age and the volume at that age.  However, it does seem to me that to avoid Boltzmann brain domination, which is apparently statistically excluded by the order within our observations that would be very improbable for Boltzmann brain observations, the variation of the weight factors with location should be nonlocal (e.g., depending on the age or time from some preferred hypersurface within the spacetime or at its past boundary).  If the weight factors depended just locally upon the conditions where the localized projection operator acts, it would seem that quantum fluctuations could reproduce these local conditions at arbitrarily late times within the universe, again leading to Boltzmann brain domination.

\section{Intrinsic Weighting for Different Observations}

Another modification that seems to be needed is not directly to use localized projection operators in the weighted integral over spacetime, but to weight the individual localized projection operators by intrinsic weight factors that depend on the efficacy of the corresponding matter configuration (e.g., brain configuration) for producing the observation.  For example, to explain why we observe ourselves to be humans despite the much greater number of ants on earth, it seems plausible to postulate that human brains are much more efficient in producing conscious perceptions than ants are, so that the total probability of human conscious observations is not far below that of ants to make human observations much less probable as a result of the much greater relative probabilities of the ant observations.  Therefore, one might postulate that the localized projection operators to human brain configurations should be given much greater weights than the localized projection operators to ant brain configurations.\\ 

Perhaps some part of this weighting factor should be the complexity of the corresponding brain, since human brains seem to be much more complex than ant brains, but I am sceptical of the hypothesis that simply getting high complexity (or high information processing) is by itself sufficient for getting a large weight multiplying the corresponding localized projection operator for the corresponding awareness operator.  But even if we do not know precisely what it is, surely there is a difference in the efficiency of different matter configurations, such as different brains or different computers, for producing different conscious observations, and these differences should be incorporated as different weights for the corresponding localized projection operators to be integrated over spacetime with the further coordinate-independent but spacetime-dependent weights such as volume averaging and Agnesi weighting.

\section{Total Weight for Each Observation}

Suppose that the expectation value of a given spacetime $S_k$ is $E_k$ and that in this spacetime the expectation value of the localized projection operator for the matter configurations corresponding to the observation $O_j$ is $n_{jk}(x^\alpha)$ as a function of the spacetime position $x^\alpha$.  Suppose further that in this spacetime the weight factor for the integration over the 4-volume is $W_k(x^\alpha)$ (e.g., $W_k(x^\alpha)= 1/[V(1+t^2)]$) and that the intrinsic weight factor for the observation $O_j$ is $w_j$.  Then the relative probability (actually, unnormalized weight) of the observation $O_j$ is
\ba
p_j = \langle A_j \rangle = w_j \sum_k E_k \int_{S_k} \sqrt{-g}\, d^4 x\, W_k(x^\alpha)\, n_{jk}(x^\alpha).
\label{b-exp}
\ea
This shows that when we break up the expectation value of the awareness operator $A_j$ corresponding to the observation $O_j$ into contributions by spacetimes $S_k$ that each have the expectation value $E_k$, one gets the intrinsic weight $w_j$ of the observation multiplied by the sum, weighted by $E_k$, of the integral over each spacetime, weighted within the spacetime $S_k$ by the location-dependent weight factor $W_k(x^\alpha)$, of the spatial density $n_{jk}(x^\alpha)$ in this spacetime that is given by the expectation value of the localized projection operator corresponding to the observation.  Therefore, in this spacetime way of proceeding, to get the unnormalized probability $p_j$ of the observation $O_j$, one needs not only the localized projection operator whose expectation value gives the expected density in spacetime of the occurrences of the matter configurations giving rise to the observation, but also the intrinsic weight of the observation, $w_j$ (e.g., the factor that is greater for human observations than for ant observations), and the weighting $W_k(x^\alpha)$ over the 4-volume of each contributing spacetime $S_k$ that renders the spacetime integral finite and avoids the potential Boltzmann brain problem.\\ 

This shows that there is a lot of freedom in the modification of Born's rule for cosmology, even within the restrictions of Sensible Quantum Mechanics in which the relative probabilities $p_j$ of observations are expectation values of the positive quantum operators $A_j$ that are the `awareness operators' corresponding to the observations $O_j$.  There would be even more freedom if one abandoned SQM and allowed the relative probabilities to be nonlinear in the expectation values $\langle A_j \rangle$, such as these expectation values to a power $s$ that is different from unity, or to be even more general nonlinear functionals of the quantum state.

\section{Comparison with Born's Rule} 

One might wonder how the freedom in the modification of Born's rule fits with arguments for Born's rule, such as the arguments of Sebens and Carroll \cite{Sebens:2014iwa,Carroll:2014mea} that the Born rule follows from their ESM-QM principle, that the probabilities an observer assigns to recorded outcomes of measurements should only depend on the joint density matrix of the observer and the detector.  In the example above, other than the weight factors and the expectation values for the spacetimes themselves, the relative probability $p_j$ of an observation depends only on the expectation values of the localized projection operators in the regions that contribute to these expectation values, which regions can be considered to be the subsystem of the observer, whose density matrix thus determines the relative probability of the observation.  However, Sebens and Carroll also make the implicit assumption that the probability of an observation is the same as the probability of the corresponding branch of the wavefunction, which puts further restrictions on the probabilities of observations beyond my assumption that they depend on the expectation values of the awareness operators.\\

One might also wonder how the freedom in the modifications of Born's rule can be consistent with the observational evidence in favor of it.  First, when one considers human observations on earth that, as observations, are confined to an extremely tiny fraction of the universe, the variation of the weight factor $W_k(x^\alpha)$ with location gives a negligible effect if this weight factor changes by order unity only over a scale comparable to that of the universe.  Second, for observations of alert humans, the intrinsic weight factors $w_j$ for different observations may well be sufficiently near each other that these differences are not readily noticed, though they might explain increased awareness of unusual events such as striking coincidences that tend to capture one's attention and plausibly lead to higher probabilities for such conscious observations.\\

If one goes beyond Sensible Quantum Mechanics, with the linear relation it gives between the quantum state and the unnormalized relative probabilities $p_j$ as the expectation values $\langle A_j \rangle$ of awareness operators, to a nonlinear relation such as $p_j = \langle A_j \rangle^s$ with $s\neq 1$, one might ask how it can be consistent with observations to have $s$ significantly different from unity.  The key is to remember that $p_j$ is the relative probability of an observation itself, which I am taking to be a conscious perception, and not the fraction of results recorded one way rather than another by some unconscious recording device.\\

For example, suppose some detector records the result of some measurement that according to Born's rule has approximately a gaussian distribution (e.g., a binomial distribution for a large number of measurements) with some standard deviation.  Assuming an idealized faithful coupling to a human brain, suppose that the relative probability of a conscious observation of the measurement result has a relative probability $p_j = \langle A_j \rangle^s$.  This would again give an approximately gaussian distribution with the same mean, but with a standard deviation $s$ times smaller than the standard deviation given by the Born rule.  Hence for $s>0$ and a sufficiently large number of measurements, the probabilities for the human conscious observations would be concentrated on the fractions of results of one kind versus another that would be close to what the Born rule predicts for the fractions with the highest probabilities, tending to confirm the Born rule for any $s$ not too small.  Of course, for $s=0$, which is analogous to counting each branch of the wavefunction equally, the probabilities for the different observations of the fractions would not be dominated by what the Born rule gives for the expectation value of the fraction, so $s=0$ is strongly statistically ruled out by observations of fractions different from a uniform distribution of results.\\

If the observer is aware not just of the total number of measurement results of each kind but also of a whole sequence of sub-results, within the single human observation (e.g., one single conscious perception) there is an awareness of the fluctuations between the sub-results.  If these fluctuations are significantly more than what would be predicted by Born's rule, this would give statistical evidence for $s<1$, and if they are significantly less than what Born's rule would predict, this would be evidence for $s>1$.  An awareness of sub-results whose fluctuations are within the range of what Born's rule would predict would give evidence against $s$ being too much smaller than 1 and also against $s$ being too much larger than 1.  However, the precision to which $s$ could be given would be limited by the number of different sub-results that one would be consciously aware of at once, say $N$, and I suspect that the statistical uncertainty of $s$ would be of the order of $1/\sqrt{N}$, so with a reasonable limit on the number of sub-results one can be simultaneously consciously aware of in a single conscious perception, I doubt that $s$ could be determined by humans to a precision of even a few percent.  Note that it would not help to have some device record the fluctuations in the sub-results and report them to the conscious observer, since the peak in the observational relative probabilities $p_j = \langle A_j \rangle^s$ for the observations of the fluctuations recorded by the device would be the same as the peak in the Born rule distribution of the fluctuations if the transfer of the information is faithful from the device to the conscious system that gives the contributions to $\langle A_j \rangle$.\\

Thus the observational support for Born's rule is actually also support for a nonlinear theory in which the relative probabilities of observations are $p_j = \langle A_j \rangle^s$ for any $s$ that is fairly close to unity.  It does not give statistical support for $s$ being extremely close to unity, though of course $s=1$ is the simplest possibility and hence by Occam's razor could be assigned a high prior probability of being precisely true, as I personally generally assume is indeed the case.  However, it is worthwhile being aware that taking $s$ to be precisely unity is just a simple assumption and, apart from its simplicity, does not seem to have strong observational support.  Therefore, considering the example of $s$ not being exactly unity is a good foil against the certainty of assumptions that are needed to prove Born's rule, though considerations of the possible variation of weight factors such as the intrinsic weight $w_j$ of an observation, and the weight factor $W_k(x^\alpha)$ for the integration over the spacetime $S_k$, which I view as much more plausible, also give counterexamples to the assumptions leading to Born's rule.

\section*{Acknowlegments}

I am grateful for discussions with Sean Carroll and for his hospitality at Caltech, where this paper was written.  He emphasized that my Sensible Quantum Mechanics framework with the relative probability of the observation $O_j$ being the expectation value of a positive awareness operator $A_j$ (or the generalization of SQM to $p_j = \langle A_j \rangle^s$) is {\it not} Everettian quantum theory, which assumes a particular branching of the quantum state and probabilities of observation depending on this branching.  This work was supported in part by the Natural Sciences and Engineering Research Council of Canada and was completed with the gracious hospitality of the Mitchell family at Cook's Branch Conservancy.





\begin{thebibliography}{}

\bibitem{Born}
  M.~Born,
  ``Zur Quantenmechanik der Sto{\ss}vorg\"{a}nge,'' 
  Z.\ Phys.\ {\bf 37}, 863-867 (1926).

\bibitem{Page:2008ns} 
  D.~N.~Page,
  ``Insufficiency of the Quantum State for Deducing Observational Probabilities,''
  Phys.\ Lett.\ B {\bf 678}, 41 (2009)
  [arXiv:0808.0722 [hep-th]].

\bibitem{Page:2009qe} 
  D.~N.~Page,
  ``The Born Rule Fails in Cosmology,''
  JCAP {\bf 0907}, 008 (2009)
  [arXiv:0903.4888 [hep-th]].

\bibitem{Page:2009mb} 
  D.~N.~Page,
  ``Born Again,''
  arXiv:0907.4152 [hep-th].

\bibitem{Page:2010bj} 
  D.~N.~Page,
  ``Born's Rule Is Insufficient in a Large Universe,''
  arXiv:1003.2419 [hep-th].

\bibitem{Page:1995dc}
  D.~N.~Page,
  ``Sensible Quantum Mechanics: Are Only Perceptions Probabilistic?,''
  quant-ph/9506010.

\bibitem{Page:1995kw} 
  D.~N.~Page,
  ``Sensible Quantum Mechanics: Are Probabilities Only in the Mind?,''
  Int.\ J.\ Mod.\ Phys.\ D {\bf 5}, 583 (1996)
  [gr-qc/9507024].

\bibitem{Page:2001ba} 
  D.~N.~Page,
  ``Mindless Sensationalism: A Quantum Framework for Consciousness,''
  in {\it Consciousness: New Philosophical Perspectives}, eds.\ Quentin Smith and
  Alexander Jokic (Oxford, Oxford University Press, 2003), pp.\ 468-506
  [quant-ph/0108039].
  
\bibitem{Page:2011sr} 
  D.~N.~Page,
  ``Consciousness and the Quantum,''
  arXiv:1102.5339 [quant-ph].

\bibitem{Page:2014aqa} 
  D.~N.~Page,
 ``Cosmological Ontology and Epistemology,''
  in {\it Philosophy of Cosmology}, eds.\ K.\ Chamcham, J.\ Silk, J.\ D.\ Barrow
  and S. Saunders (Cambridge University Press, Cambridge, 2017), pp.\ 317-329
  \newline
  [arXiv:1412.7544 [physics.hist-ph]].
  
\bibitem{DKS}
  L.~Dyson, M.~Kleban and L.~Susskind,
  ``Disturbing Implications of a Cosmological Constant,''
  JHEP {\bf 0210}, 011 (2002) [hep-th/0208013].

\bibitem{Albrecht:2004}
  A.~Albrecht,
  ``Cosmic Inflation and the Arrow of Time,''
  in {\it Science and Ultimate Reality: Quantum Theory, Cosmology, and Complexity},
  edited by J.~D.~Barrow, P.~C.~W.~Davies, and C.~L.~Harper, Jr.\ 
  (Cambridge University Press, Cambridge, 2004), pp. 363-401, [astro-ph/0210527].

\bibitem{Page:2008a}
  D.~N.~Page,
  ``Is Our Universe Likely to Decay within 20 Billion Years?,''
  Phys.\ Rev.\ D {\bf 78}, 063535 (2008) [hep-th/0610079].
  
\bibitem{Page:2006ys}
  D.~N.~Page,
  ``Return of the Boltzmann Brains,''
  Phys.\ Rev.\ D {\bf 78}, 063536 (2008) [arXiv:hep-th/0611158 [hep-th]].

\bibitem{Page:2008b}
  D.~N.~Page,
  ``Is Our Universe Decaying at an Astronomical Rate?,''
  Phys.\ Lett.\ B {\bf 669}, 197 (2008) [hep-th/0612137].
  
\bibitem{Page:2017qdu}
D.~N.~Page,
``Bayes Keeps Boltzmann Brains at Bay,''
[arXiv:1708.00449 [hep-th]].  
  
\bibitem{Page:2008zh} 
  D.~N.~Page,
  ``Cosmological Measures without Volume Weighting,''
  JCAP {\bf 0810}, 025 (2008)
  [arXiv:0808.0351 [hep-th]].

\bibitem{Page:2011gq} 
  D.~N.~Page,
  ``Cosmological Measures with Volume Averaging,''
  Int.\ J.\ Mod.\ Phys.\ Conf.\ Ser.\  {\bf 1}, 80 (2011).

\bibitem{Page:2010re} 
  D.~N.~Page,
  ``Agnesi Weighting for the Measure Problem of Cosmology,''
  JCAP {\bf 1103}, 031 (2011)
  [arXiv:1011.4932 [hep-th]].

\bibitem{Page:2014eoa} 
  D.~N.~Page,
  ``Spacetime Average Density (SAD) Cosmological Measures,''
  JCAP {\bf 1411}, no. 11, 038 (2014)
  [arXiv:1406.0504 [hep-th]].

\bibitem{Sebens:2014iwa} 
  C.~T.~Sebens and S.~M.~Carroll,
  ``Self-Locating Uncertainty and the Origin of Probability in Everettian Quantum Mechanics,''
  The British Journal for the Philosophy of Science {\bf 69} (1), 25-74 (2018).   
  [arXiv:1405.7577 [quant-ph]].

\bibitem{Carroll:2014mea} 
  S.~M.~Carroll and C.~T.~Sebens,
  ``Many Worlds, the Born Rule, and Self-Locating Uncertainty,''
  in {\it Quantum Theory: A Two-Time Success Story, Yakir Aharonov Festschrift},
  eds.\ D.\ C.\ Struppa and J.\ M.\ Tollaksen (Springer-Verlag, 2013) p. 157.
  [arXiv:1405.7907 [gr-qc]].

\end{thebibliography}
\end{document}